\documentclass[lettersize,journal]{IEEEtran}
\usepackage{graphicx}
\usepackage{subcaption}
\usepackage{mathtools}
\usepackage{amsmath,amsfonts}
\usepackage{amssymb}
\usepackage{color,soul}
\usepackage{algorithm}
\usepackage{algpseudocode}
\usepackage{array}
\usepackage[caption=false,font=normalsize,labelfont=sf,textfont=sf]{subfig}
\usepackage{textcomp}
\usepackage{mathtools}
\usepackage{stfloats}
\usepackage{url}
\usepackage{verbatim}
\usepackage[acronym,toc]{glossaries}
\usepackage{lipsum}
\usepackage{balance}
\usepackage{cite}
\usepackage{booktabs}

\newcommand{\bigO}[0]{\mathcal{O}}

\def\BibTeX{{\rm B\kern-.05em{\sc i\kern-.025em b}\kern-.08em
    T\kern-.1667em\lower.7ex\hbox{E}\kern-.125emX}}

\newcommand{\etal}[0]{\textit{et al} }
\newcommand\numberthis{\addtocounter{equation}{1}\tag{\theequation}}

\date{April 2025}
\DeclareMathOperator*{\minimize}{minimize}
\DeclareMathOperator*{\subto}{subject\;to}
\DeclareMathOperator{\tr}{tr}
\DeclareMathOperator{\herm}{^{\mathsf{H}}}
\DeclareMathOperator{\vectorial}{vec}
\newacronym{CS}{CS}{compressed sensing}
\newacronym{CSI}{CSI}{channel state information}
\newacronym{AUD}{AUD}{active users detection}
\newacronym{NOMA}{NOMA}{non-orthogonal multiple access}
\newacronym{MP}{MP}{message passing}
\newacronym{AMP}{AMP}{approximate message passing}
\newacronym{BS}{BS}{base station}
\newacronym{MTC}{MTC}{machine type communications}
\newacronym{SNR}{SNR}{signal-to-noise-ratio}
\newacronym{LoS}{LoS}{line of sight}
\newacronym{NLoS}{NLoS}{non-LoS}
\newacronym{OMP}{OMP}{orthogonal matching pursuit}
\newacronym{ADMM}{ADMM}{alternating direction method of multipliers}
\newacronym{ADMM-ILI}{ADMM-ILI}{ADMM with imperfect location information}
\newacronym{ADMM-LI}{ADMM-LI}{ADMM with perfect location information}
\newacronym{TS}{TS}{time slot}
\newacronym{IoT}{IoT}{Internet of Things}
\newacronym{RIS}{RIS}{reconfigurable intelligent surfaces}
\newacronym{MLE}{MLE}{maximum likelihood estimation}
\begin{document}

\title{Active IoT User Detection in Near-Field with Location Information}
\author{Gabriel Martins de Jesus, Richard Demo Souza, Onel Luis Alcaraz L{\'o}pez
\thanks{G. M. de Jesus and O. L. A. L{\'o}pez are with the Centre for Wireless Communications (CWC), University of Oulu, 90014 Oulu, Finland. (e-mail: gabriel.martinsdejesus@oulu.fi, onel.alcarazlopez@oulu.fi).}
\thanks{R. D. Souza is with the Electrical and Electronics Engineering Department, Federal University of Santa Catarina (UFSC), Florian{\'o}polis 88040-900, Brazil (e-mail: richard.demo@ufsc.br).}
\thanks{This research has been supported by the Research Council of Finland (former Academy of Finland) Grant 346208 (6G Flagship Programme) and Grant 362782 (ECO-LITE), CNPq (403124/2023-9, 305021/2021-4) and RNP/MCTI Brasil 6G project (01245.020548/2021-07).}}

\markboth{Journal of \LaTeX\ Class Files,~Vol.~18, No.~9, September~2020}%
{How to Use the IEEEtran \LaTeX \ Templates}

\maketitle

\begin{abstract}
    In this paper, we address active users detection (AUD) in near-field Internet of Things (IoT) networks by exploring prior knowledge of users' locations. We consider a scenario where users are distributed in a semi-circular area within the Rayleigh distance of a multi-antenna base station (BS). We propose the BS to use location estimates of the users to reconstruct their line-of-sight (LoS) channel components, hence assisting the AUD process. For this, the BS combines these reconstructed channels with users' pilot sequences, enhancing the correlation between received signals and active users. We formulate the location-aided AUD as a convex optimization problem, solved via the alternating direction method of multipliers (ADMM). {Our proposal has a higher computational complexity compared to the baseline ADMM approach where location information is not used. Moreover, the proposal requires location information of users, which can be readily informed if users are static, or inferred via established localization algorithms if they are mobile.} Simulation results compare our proposal against the baseline across varying systems parameters, such as number of users, pilot length and LoS component strength. We demonstrate that under perfect location estimation and strong LoS, our proposed method significantly outperforms the baseline. Furthermore, robustness analysis shows that performance gains persist under imperfect location estimation, provided the estimation error remains within bounds determined by the system parameters.
\end{abstract}


\section{Introduction}

\IEEEPARstart{E}{lectronic} devices performing \gls{MTC} are becoming predominant within modern wireless networks \cite{Shariatmadari:2015:ICM}. In future 6G networks, \gls{MTC} will likely be dominated by non-critical \gls{IoT} applications, characterized by devices with limited computational capabilities and strict low-power requirements \cite{Braud:2021:book}. To accommodate these constraints, grant-free random access schemes are employed to minimize signaling overhead between the \gls{BS} and users, significantly reducing power consumption. {While this approach increases the probability of packet collisions, potentially leading to transmission failures, it is specially well suited for the sparse traffic nature of \mbox{\gls{IoT}} devices.} Since \gls{IoT} devices are typically configured to transmit only when new data is acquired, they exhibit sporadic activity and are rarely active simultaneously \cite{Marata:2023:ITWC}. This inherent sparsity in \gls{MTC} networks is a critical property exploited during the \gls{AUD} phase.

In the \gls{AUD} phase, the \gls{BS} aims to detect and identify the specific subset of active users from the total device pool \cite{Lopez:2023:PotI}. Users are assigned pilot sequences which are attached as preambles to their information packets, and these are utilized to identify active users and estimate their respective channels based on the received signal. When multiple users transmit simultaneously, their pilots arrive at the receiver superimposed and distorted by the wireless channel, making direct user separation challenging for the \gls{BS}. However, due to the high sparsity of \gls{MTC} traffic, established detection methods can be used for \gls{AUD}, including \gls{CS} \cite{Knoop:2014:EUSIPCO}, machine learning \cite{Kim:2020:ITC}, and convex optimization-based algorithms such as \gls{ADMM} \cite{Djelouat:ITWC:2022}. Accurate \gls{AUD} is a fundamental prerequisite for reliable channel estimation and subsequent data detection \cite{Bockelmann:2016:ICM}.

Looking beyond 5G, the transition to near-field communications is a promising scenario for 6G technologies \cite{Cui:2023:ICM}. As infrastructure evolves toward increasingly large antenna arrays and higher frequency bands, the near-field region extends significantly, ranging from a few centimeters to hundreds of meters \cite{Liu:2023:OJCOMS}. Within this region, the classical planar wave assumptions of far-field models are invalid, and spherical wavefronts are considered instead\cite{Lu:ICST:2024}. Unlike planar models, the spherical wave model captures phase variations across the array that depend on both the angle of arrival and the specific distance of the source. Due to these propagation characteristics, techniques such as beam-focusing in the downlink \cite{Lu:ICST:2024} and user localization based solely on uplink signals \cite{Chen:2024:IWC, Haghshenas:arxiv:2025, Gurgunoglu:IWCL:2025} are possible. The latter is crucial for this work, and this localization capability can be explored to enhance the \gls{AUD} problem, as by determining a user's position, the system can reconstruct the deterministic \gls{LoS} channel response, which typically serves as the dominant signal component. This \gls{LoS} component, in turn, can be combined with the pilot sequence to better differentiate between users, improving the accuracy of the \gls{AUD}.

\subsection{Related work}
\gls{AUD} is typically paired with channel estimation, and solved jointly. To achieve better performance in detection accuracy, many works proposed modifications to established \gls{CS} algorithms  \cite{deJesus:2024:GLOBECOM,Djelouat:2021:Asilomar,Marata:ITWC:2024,Wang:2024:VTC,Zhang:2025:IOTJ}, such as the \gls{OMP} \cite{Donoho:2006:SP} and \gls{AMP} \cite{Maleki:2010:CISS}. In each of these works, the proposed modifications take into account specific system model assumptions, such as detection of the same user in different frequency channels \cite{deJesus:2024:GLOBECOM}, spatially correlated channels \cite{Djelouat:2021:Asilomar}, users clustered by non-orthogonal pilot sequences \cite{Marata:ITWC:2024}, as well as including protection from malicious attacks via pilot contamination \cite{Wang:2024:VTC} and specifically tailor them for satellite networks \cite{Zhang:2025:IOTJ}. \gls{AUD} has also been solved via Bayesian learning methods \cite{Zhang:2021:ITVT} and neural networks \cite{Sivalingam:2021:PIMRC,Ahn:2022:IOTJ,Sivalingam:2025:IOTJ,Sun:2025:IOTJ}. Interestingly, although the typical formulation of \gls{AUD} is a non-convex problem, it can be relaxed to be solved with convex optimization algorithms \cite{Djelouat:ITWC:2022,Afshar:2022:ISPL,Liu:2023:GLOBECOM,Torabnezhad:2024:TS}. Meanwhile, if the channels have been previously estimated, this information can be used to improve the \gls{AUD} and assist the data detection. For instance, in a multi-carrier setup, Yang \etal \cite{Yang:2024:ITWC} do not consider transmission of pilot sequences, but instead rely on the \gls{CSI} and knowledge of the spread sequences to estimate which users are active. The initial detection is made based on an energy test of the received signal in each carrier, declaring active the users that transmit in the carriers that are not idle. Then, a false-alarm mechanism is proposed, iteratively removing users based on the symbols detected by the \gls{AMP}-based data detector, significantly decreasing false-alarm ratio and symbol error rate.

In the far-field, the \gls{AUD} problem has been extensively explored, while its study in the near-field has increased recently \cite{Wang:2025:ITWC,Mylonopoulos:2025:IWCL,Arai:2025:ITWC,Qiao:2024:ITC,Zhang:2025:INFOCOM}.
Wang \etal \cite{Wang:2025:ITWC} model the near-field channel with correlated Rician fading. They formulate \gls{AUD} as a \gls{MLE} problem, which is significantly more complicated than with uncorrelated channels. The performance of the proposal is evaluated against the typical uncorrelated channels assumptions, outperforming the latter as the total number of users increase or the pilot length decreases.
Mylonopoulos \etal \cite{Mylonopoulos:2025:IWCL} proposed a technique that jointly detects the active users and estimates their positions with the aid of several \gls{RIS}. Here, users are not assigned individual pilot sequences a priori, risking collisions. Each \gls{RIS} is designed to inspect a specific area, attempting to identify the spatial origin of the incoming signals. Then, resource blocks are assigned to the users in those areas, and multipath components are suppressed by the \gls{RIS} and do not reach the \gls{BS}. Compared to the case without \gls{RIS}, the proposal improves performance when there are \gls{NLoS} components as the symbols' power increases.
Arai \etal \cite{Arai:2025:ITWC} split \gls{AUD} in two phases based on \gls{OMP}, first estimating the channels, then detecting the users and associating the channels. In the former, an \gls{OMP} algorithm is ran over a polar grid covering the deployment area of the users. Then, the detected channels are associated with the users. The proposal guarantees lower normalized mean-squared error and bit error rate than other approaches. 
Qiao \etal \cite{Qiao:2024:ITC} use an extension of \gls{OMP} to perform \gls{AUD} and channel estimation, and also propose an algorithm to estimate the position of the active users. The channel estimate of the active users is used with the MUSIC algorithm to estimate the angle of arrival and time difference of arrival and obtain the coordinates of the users with the least squares algorithm. \gls{AUD} and channel estimate performance is superior to other baselines, and the position error decreases with higher transmit powers.
Zhang \etal \cite{Zhang:2025:INFOCOM} explore the spatial stationary of the near-field by modeling the activity matrix as a three-level sparse matrix, involving the active state, location of users, and visibility regions. They propose two algorithms that do channel estimation and \gls{AUD}, either jointly or separately, outperforming the baselines with the cost of increased computational complexity.

While prior works in the far-field have utilized channel knowledge to improve \gls{AUD} performance \cite{Yang:2024:ITWC}, to the best of our knowledge, this approach has not yet been explored in the near-field context. To address this gap, our paper makes the following main contributions: \begin{enumerate} \item We formulate  \gls{AUD} as a convex optimization problem that explicitly incorporates prior knowledge of users locations. While \cite{Haghshenas:arxiv:2025,Gurgunoglu:IWCL:2025} have demonstrated that user location can be obtained in the near-field with reasonable accuracy, and \cite{Mylonopoulos:2025:IWCL,Arai:2025:ITWC,Qiao:2024:ITC,Zhang:2025:INFOCOM} obtain estimates of the location of users as a consequence of \gls{AUD}, none use location information as prior information to improve \gls{AUD} accuracy. \item We solve the \gls{AUD} with location information via \gls{ADMM}. Using \gls{ADMM} to solve the \gls{AUD} has been explored in the literature \cite{Cirik:2018:IWCL, Djelouat:2021:Asilomar, Djelouat:ITWC:2022}, but explicitly using the channel matrix adds more complexity to the formulation, and we derive the closed form expressions for obtaining the set of active users. \item We show that the proposal retains its superiority even with imperfect estimates, provided the estimation error remains below a specific network-dependent threshold. Moreover, our proposal outperforms the baseline in scenarios with a larger number of total users, active users, and antenna elements, whereas the baseline is favored by longer pilot sequences when location information errors are high. Furthermore, our results indicate that while the baseline is preferable in pure \gls{NLoS} channels, our proposal achieves superior performance provided that the \gls{LoS} component between the users and the \gls{BS} is sufficiently strong. \end{enumerate}

The remaining of this paper is organized as follows. In Section \ref{sec:system_model}, we introduce our system model, detailing all assumptions considered in this work. In Section \ref{sec:AUD_ADMM}, we introduce the convex optimization problem and present the solutions via \gls{ADMM} for the cases with known and unknown users location. In Section \ref{sec:results}, we evaluate the performance of \gls{ADMM-LI} against the baseline without location information under different network and estimation parameters. Lastly, we conclude the paper in Section~\ref{sec:conclusion}.

{Throughout this paper, we adopt the following notation. Matrices are denoted by $\mathbf{A}$, and vectors by $\mathbf{a}$. For matrix $\mathbf{B}$, its $i$-th row (or column) is denoted by $\mathbf{b}_i$, and the element at the $i$-th row and $j$-th column is denoted by $B_{i,j}$. For matrix $\mathbf{A}$, $||\mathbf{A}||_F$ denotes its Frobenius. The $\text{diag}(\mathbf{c})$ operator takes vector $\mathbf{c}$ and transforms it into the matrix $\mathbf{C}$, with each diagonal element given by the elements in $\mathbf{c}$, and off-diagonal elements equal to 0, while the  $\text{vec}(\mathbf{D})$ operator takes the $P\times Q$ matrix $\mathbf{D}$ and flattens it into the $1 \times PQ$ vector $\mathbf{d} = [\mathbf{d}_1, \mathbf{d}_2, \dots, \mathbf{d}_P]$}.

\section{System Model}\label{sec:system_model}
\begin{figure}
    \centering
    \includegraphics[width=\linewidth]{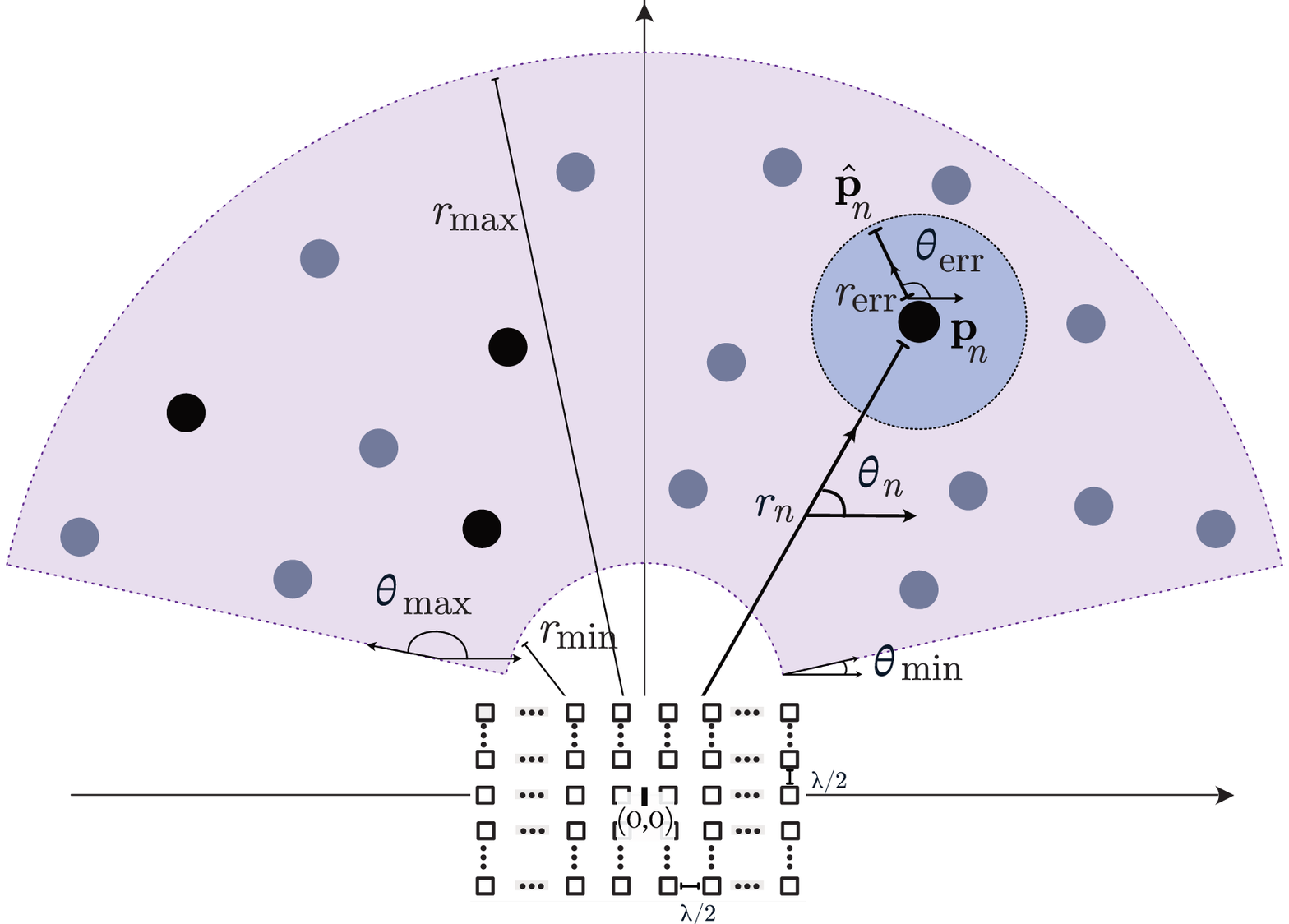}
    \caption{Illustration of the system model, where users are indicated by the filled circles, with black filling indicating the $K$ active users.}
    \label{fig:system_model}
\end{figure}

We consider an \gls{IoT} network with $N$ single-antenna devices served by a single \gls{BS} that is centered at the origin, as illustrated in Fig. \ref{fig:system_model}. The distance of the $n$-th user to the origin is denoted by ${r}_n$, and its angle to the X-axis is denoted by $\theta_n$. The users are placed in a semi-circular area, and are within the region limited by $r_{\text{min}} \leq r_n \leq r_{\text{max}}$, and $\theta_{\text{min}} \leq \theta_n \leq \theta_{\text{max}}$. In cartesian coordinates, $\mathbf{p}_n=[r_n\cos{\theta_n}, r_n\sin{\theta_n}]$ denotes the position of the $n$-th user. The \gls{BS} has an estimate of the position of users, but errors in the estimate are present in practical deployments, which we model here by introducing the random variables $r_{\text{err}}\sim \mathcal{N}(0,\sigma)$ and $\theta_{\text{err}}\sim\mathcal{U}(0,2\pi)$ that are combined and summed to the true position of the users. The resulting is a position estimate $\hat{\mathbf{p}}_n$ that is in a point in the circumference of radius $r_{\text{err}}$ and center $\mathbf{p}_n$, i.e.,
\begin{equation}
    \hat{\mathbf{p}}_n = \mathbf{p}_n + [r_{\text{err}}\cos\theta_{\text{err}}, r_{\text{err}}\sin \theta_{\text{err}}].
\end{equation}

At each \gls{TS}, any user may generate a packet with probability $p=K/N$, and transmit it in the shared frequency channel. We denote by $\mathcal{K}$ the set of active users. The \gls{BS} is equipped with an antenna array of $M$ elements centered at the origin and with each element $m$ located in $\boldsymbol{\delta}_m$. Each element is spaced from the others by distance of at least $d=\lambda/2$, where $\lambda=c/f$ is the wavelength of the carrier, and $c$ is the speed of light. 

\subsection{Channel model}
We consider that the users are in the radiative field and their distance is below the Rayleigh distance $d_R \triangleq 2D^2/\lambda$, i.e., ~$0.62\sqrt{D^3/\lambda}\leq r_{n} \leq d_R$ \cite{Lu:ICST:2024}, where $D$ denotes the dimension of the antenna array and depends on its topology, as well as number of antenna elements. In such region, users experience near-field effects. {In the near-field, if the antenna elements are sufficiently close, the magnitude of the received power of a given signal can be modeled as equal in all $M$ elements.} However, the phase is nonlinear and dependent on the distance from the user to each element \cite{Lu:ICST:2024}. Let $h^{\text{LoS}}_{n,m}$ denote the \gls{LoS} channel between the $n$-th user and the $m$-th antenna element, such that
\begin{equation}\label{eq:hnm}
    h^{\text{LoS}}_{n,m}= \alpha_n e^{j\varphi_{n,m}},
\end{equation}
with
\begin{equation}
    \alpha_n = \frac{\lambda}{4\pi r_n},
\end{equation}
and
\begin{equation}\label{eq:theta}
    \varphi_{n,m} = -\frac{2\pi}{\lambda}r_{n,m},
\end{equation}
where the term $r_{n,m} = ||\mathbf{p}_n - \boldsymbol{\delta}_m||_2$ is the distance between the $n$-th user and the $m$-th antenna element in the array.  The channel response is arranged in a channel vector as
$\mathbf{h}^{\text{LoS}}_n = \{h^{\text{LoS}}_{n,m}\}_{m=1}^M$, and the channel matrix is defined as 
\begin{equation}\label{eq:H}
    \mathbf{H}^{\text{LoS}} = [\mathbf{h}^{\text{LoS}}_1,\mathbf{h}^{\text{LoS}}_2,\dots,\mathbf{h}^{\text{LoS}}_N]^\intercal\in \mathbb{C}^{N\times M}.
\end{equation}
Moreover, the \gls{NLoS} components follow a complex Gaussian normal distribution, with each element denoted by $h^{\text{NLoS}}_{n,m}$, and arranged in the channel vector $\mathbf{h}^{\text{NLoS}}_n$. The vectors, in turn, are arranged as $\mathbf{H}^{\text{NLoS}} = [\mathbf{h}^{\text{NLoS}}_{1},\mathbf{h}^{\text{NLoS}}_{2},\dots,\mathbf{h}^{\text{NLoS}}_{N}]^\intercal\in \mathbb{C}^{N\times M}$. We consider a Ricean channel with parameter $\mu$, which relates the total power of the \gls{LoS} to the \gls{NLoS} components. The resulting channel is given by
\begin{equation}\label{eq:H_tot}
    \mathbf{H}^{\text{total}} = \sqrt{\frac{\mu}{1+\mu}}\mathbf{H}^{\text{LoS}} + \sqrt{\frac{1}{1+\mu}}\mathbf{H}^{\text{NLoS}}.
\end{equation}

\subsection{Signal model}
Before transmitting data, an active user will transmit its pilot sequence to the \gls{BS}. The \gls{BS} is aware of all sequences, which are unique to each user, and are used to identify the active users and potentially estimate their channels. We denoted by $T$ the total number of symbols allocated to transmitting the pilot sequence, and we assume that $T$ is smaller than the coherence interval in the time frame. Let $\pmb{\phi}_n\in\mathbb{C}^{T\times 1}$ denote the pilot sequence assigned to the $n$-th user, with $||\pmb{\phi}_n||_2=1$. Then, the pilot sequence matrix is defined as $\boldsymbol{\Phi}\triangleq[\pmb{\phi}_1, \pmb{\phi}_2, \dots, \pmb{\phi}_N]\in\mathbb{C}^{T\times N}$. Moreover, we consider $T\leq N$, such that, although unique, the users' pilot sequences are not pair-wise orthogonal, which results in \textit{soft} collisions. Unlike in \textit{hard} collisions, where more than a single user transmits the same pilot sequence, it is still possible to separate the incoming signals when soft collisions occur.

We denote by $\mathbf{X}$ the activity matrix, defined as {$\text{diag}([x_1, x_2, \dots, x_N])$, with $x_{n}=\sqrt{\gamma_n}$ if $n\in \mathcal{K}$, and $x_n=0$ otherwise.} The users use inverse power control based on their location information, such that, $\gamma_n = \left({4\pi||\hat{\mathbf{p}}_n||_2}/{\lambda}\right)^2$ is the transmit power of user $n$. This leads to the power of the received signal to be unitary if there was only \gls{LoS} and if $\hat{\mathbf{p}}_n={\mathbf{p}}_n${\footnote{{Although other power control approaches (or no power control at all) could be considered without affecting the development of the proposal, we set for this specifically to avoid the risk of the algorithm being biased towards users closer to the {\gls{BS}}.}}}. 

The received signal $\mathbf{Y}\in\mathbb{C}^{T\times M}$ is then given by
\begin{equation}\label{eq:mine_received}
    \mathbf{Y} = \boldsymbol{\Phi}\mathbf{X}\mathbf{H}^\text{total} + \mathbf{V},
\end{equation}
where $\mathbf{V}\in\mathbb{C}^{T\times M}$ is the additive white Gaussian noise.   
Another way to write the received signal is by defining $\mathbf{J}\triangleq(\mathbf{XH}^\text{total})\in\mathbb{C}^{N\times M}$, which is a convenient representation when doing joint active user detection and channel estimation. In this case, the received signal is written as 
\begin{equation}\label{eq:typical_received}
    \mathbf{Y} = \boldsymbol{\Phi}\mathbf{J} + \mathbf{V}.
\end{equation}

\section{AUD with and without Location Information}\label{sec:AUD_ADMM}
During the \gls{AUD} phase, the \gls{BS} attempts to identify the subset $\mathcal{K}$ of active users in the current \gls{TS}, utilizing prior knowledge of the pilot sequences of the users pool. In grant-free access, \gls{AUD} is fundamental to identify which users are transmitting, since the typical \textit{handshake} between \gls{BS} and users does not happen. For this reason, improving the accuracy of \gls{AUD} is important, as the information packets from undetected users are lost, and noise can be interpreted to be real data from inactive users. If the \gls{LoS} component is dominant in the received signal, it can be explored to improve accuracy in the \gls{AUD} process, as the received signal will be correlated with both pilot sequence and the \gls{LoS} channel of any user active. If the \gls{BS} knows the location of the users, either by estimating it, e.g., with MUSIC, or because users are fixed and their location is informed to the network, it can reconstruct an estimate of the \gls{LoS} component. Then, these estimated components can be combined with the pilot sequences to better differentiate between users. In this section, we propose improving \gls{AUD} by explicitly incorporating location information, formulated as a convex problem and solved using \gls{ADMM}. For comparison, we also present the standard \gls{ADMM}-based \gls{AUD} formulation without location data, as detailed in \cite{Djelouat:ITWC:2022}. Throughout the paper, we refer to our proposal as \gls{ADMM-LI}, and to the formulation without location information as the baseline. 

\subsection{Problem formulation}
To formulate our problem, we assume the \gls{BS} is aware of the location of users and constructs the matrix $\hat{\mathbf{H}}$ with each entry according to \eqref{eq:hnm}, but with $\alpha_n=1$. By ignoring the \gls{NLoS} components, we model the \gls{AUD} problem as
\begin{equation}\label{eq:inicial2}
\begin{aligned}
\minimize_{\mathbf{X}} \quad & \frac{1}{2}||\boldsymbol{\Phi}\mathbf{X}\hat{\mathbf{H}} - \mathbf{Y}||_F^2 + \beta ||\mathbf{X}||_{2,0},
\end{aligned}
\end{equation}
where $\beta$ is a regularization parameter to either emphasize the sparsity of the estimated $\mathbf{X}$ or the accuracy of $\boldsymbol{\Phi}\mathbf{X}\hat{\mathbf{H}} - \mathbf{Y}$,  and  $||\mathbf{X}||_{2,0}\triangleq\sum_{n=1}^Ni(||\mathbf{x}_n||_2)$, with $i(a)$ being the indicator function with value $0$ if $a=0$ and $1$ otherwise.
The problem in \eqref{eq:inicial2} is not convex because of the term $||\mathbf{X}||_{2,0}$. As in \cite{Djelouat:ITWC:2022}, the objective function needs to be relaxed. While we can relax the $\ell$-0 norm to $\ell$-1, it will be biased towards entries in $\mathbf{X}$ with high magnitude. An alternative is the log-sum penalty, resulting in $\sum_{n=1}^N \log{(u_n + \epsilon_0)},$
where $||\mathbf{x}_n||_2 \leq u_n, n=\{1, 2, \dots, N\}$, and $\mathbf{x}_n$ and $u_n$ denote the $n$-th row and entry of $\mathbf{X}$ and $\mathbf{u}$, respectively. Then, we consider the majority-minimization approximation, obtaining $\sum_{n=1}^N \nu_n^{(r)}||\mathbf{x}_n||_2$, with $\nu_n^{(r)} =  (\epsilon_0 + ||\mathbf{x}_n||_2)^{-1}$. The relaxed problem is written as
\begin{equation}\label{eq:problem}
\begin{aligned}
    \minimize_\mathbf{X}   \frac{1}{2}||\boldsymbol{\Phi}\mathbf{X}\hat{\mathbf{H}} - \mathbf{Y}||_F^2 + \beta\sum_{n=1}^N \nu_n^{(r)}||\mathbf{x}_n||_2,
\end{aligned}
\end{equation}
{and $\nu_n^{(r)}$ is updated after solving $\mathbf{X}$ each time, until the $R$-th iteration.}

\subsection{Solution via ADMM}
The \gls{ADMM} is an algorithm to solve convex optimization problems, including \gls{AUD} \cite{Cirik:2018:IWCL, Djelouat:2021:Asilomar, Djelouat:ITWC:2022}, that combines the separability of dual ascent methods and the convergence properties of the method of multipliers \cite{Boyd:book:2011}. \gls{ADMM} requires that the objective function is separable, which is the case in \eqref{eq:problem}. To apply it, we split $\mathbf{X}$ into $\mathbf{X}$ and $\mathbf{Z}$, and write the equivalent problem as
\begin{align*}\label{eq:XZ_info}
    &\minimize_{\mathbf{X},\mathbf{Z}} \frac{1}{2}||\boldsymbol{\Phi}\mathbf{Z}\hat{\mathbf{H}} - \mathbf{Y}||_F^2 + \beta\sum_{n=1}^N \nu_n||\mathbf{x}_n||_2 \\
    \quad & \subto \quad \mathbf{X} = \mathbf{Z}. \numberthis
\end{align*}    
We define the Lagrangian dual variable $\mathbf{W}\in\mathbb{C}^{N\times N}$ and obtain the augmented Lagrangian of \eqref{eq:XZ_info} as
\begin{align*}\label{eq:Lagrangian}
&\mathcal{L}(\mathbf{X},\mathbf{Z};\mathbf{W})=\frac{1}{2}||\boldsymbol{\Phi}\mathbf{Z}\hat{\mathbf{H}} - \mathbf{Y}||_F^2 + \beta \sum_{n=1}^N \nu_n^{(r)}||\mathbf{x}_n||_2 \\ &+ \frac{\rho}{2}||\mathbf{X}-\mathbf{Z}+{\mathbf{W}}/{\rho}||_F^2 + \frac{||\mathbf{W}||_F^2}{2\rho},\numberthis
\end{align*}
where $\rho>0$ is a penalty parameter. Then, the problem is solved by iteratively minimizing the gradient of $\mathcal{L}(\mathbf{X},\mathbf{Z};\mathbf{W})$ for $\mathbf{Z}$, then $\mathbf{X}$, and lastly updating the dual variable $\mathbf{W}$ with $\rho$ as the step size. At each iteration $s$, until $s=S$, the variables are solved and updated as
\begin{equation}\label{eq:Z_info_final}
\begin{split}
\vectorial(\mathbf{Z}^{(s+1)}) = & (\hat{\mathbf{H}}\hat{\mathbf{H}}\herm\otimes \boldsymbol{\Phi}\herm\boldsymbol{\Phi} + \rho \mathbf{I}_{N^2})^{-1}\\ &\vectorial(\hat{\mathbf{H}}\herm\mathbf{Y}\boldsymbol{\Phi}\herm + \rho \mathbf{X}^{(s)} + {\mathbf{W}^{(s)}}),
\end{split}
\end{equation}
\begin{equation}\label{eq:X_info_final}
X_{n,n}^{(s+1)} = \frac{d_n}{|d_n|} \max\left(0, |d_n| - \beta\frac{ \nu_n^{(r)}}{\rho}\right),
\end{equation}
for $n\in\{1,2,\dots,N\}$, with ${d}_n = {Z}_{n,n}^{(s+1)}-{W_{n,n}^{(s)}}/{\rho}$
and
\begin{equation}\label{eq:W_info_final}
    \mathbf{W}^{(s+1)} = \mathbf{W}^{(s)} + \rho(\mathbf{X}^{(s+1)} - \mathbf{Z}^{(s+1)}),
\end{equation}
as it is detailed in the Appendix.

\subsection{Case without location information}
When no location information is present, the \gls{AUD} is solved by formulating the problem as \cite{Djelouat:ITWC:2022}
\begin{align*}\label{eq:inicial}
    \minimize_\mathbf{J} \frac{1}{2}||\Phi\mathbf{J}-\mathbf{Y}||_F^2 + \beta||\mathbf{J}||_{2,0},
\end{align*}
also relaxing and using the majority-minimization approximation, obtaining
\begin{equation}\label{eq:l-X}
\begin{aligned}
    \minimize_\mathbf{J}   \frac{1}{2}||\boldsymbol{\Phi}\mathbf{J} - \mathbf{Y}||_F^2 + \beta\sum_{n=1}^N \nu_n^{(r)}||\mathbf{j}_n||_2,
\end{aligned}
\end{equation}
with $\nu_n^{(r)} =  (\epsilon_0 + ||\mathbf{j}_n||_2)^{-1}$. To apply \gls{ADMM}, the problem is rewritten as 
\begin{align*}
    &\minimize_{\mathbf{J},\mathbf{Z}} \frac{1}{2}||\boldsymbol{\Phi}\mathbf{Z} - \mathbf{Y}||_F^2 + \beta \sum_{n=1}^N \nu^{(r)}_n||\mathbf{j}_n||_2 \\
    \quad & \subto \quad \mathbf{J} = \mathbf{Z},\numberthis \label{eq:JZ_noinfo}
\end{align*}    
with the Lagrangian given by
\begin{align*}
&\mathcal{L}(\mathbf{J},\mathbf{Z};\mathbf{W})=\frac{1}{2}||\boldsymbol{\Phi}\mathbf{Z} - \mathbf{Y}||_F^2 + \beta \sum_{n=1}^N \nu_n^{(r)}||\mathbf{j}_n||_2 \\ &+ \frac{\rho}{2}||\mathbf{J}-\mathbf{Z}+{\mathbf{W}}/{\rho}||_F^2 + \frac{||\mathbf{W}||_F^2}{2\rho},\numberthis
\end{align*}and the update rules for $\mathbf{J}$, $\mathbf{Z}$ and $\mathbf{W}$ are given by \cite{Djelouat:ITWC:2022}
\begin{equation}\label{eq:Z_noinfo_final}
    \mathbf{Z}^{(s+1)} = (\boldsymbol{\Phi}\herm\boldsymbol{\Phi} + \rho \mathbf{I}_N)^{-1}(\boldsymbol{\Phi}\herm\mathbf{Y} + \rho \mathbf{J}^{(s)} + \mathbf{W}^{(s)}),
\end{equation}
\begin{equation}\label{eq:j_noinfo_final}
\mathbf{j}_n^{(s+1)} = \frac{\mathbf{d}_n}{||\mathbf{d}_n||_2} \max\left(0, ||\mathbf{d}_n||_2 - \beta\frac{ \nu_n^{(r)}}{\rho}\right),
\end{equation}
for $n\in\{1,2,\dots,N\}$, with $\mathbf{d}_n = \mathbf{z}_n^{(s+1)}-{\mathbf{w}_n^{(s)}}/{\rho}$, and
\begin{equation}\label{eq:W_noinfo_final}
    \mathbf{W}^{(s+1)} = \mathbf{W}^{(s)} + \rho(\mathbf{J}^{(s+1)} - \mathbf{Z}^{(s+1)}).
\end{equation}
In Algorithm \ref{alg}, we summarize the steps for each approach.
\begin{algorithm}[tb]
\caption{\gls{ADMM} for \gls{AUD}}\label{alg}
 \hspace*{\algorithmicindent} \textbf{Input:} $\boldsymbol{\Phi}$, $\mathbf{Y}$, $\mathbf{H}$ (ADMM-LI), $S$, $R$\\
 \hspace*{\algorithmicindent} \textbf{Output:}  $\mathbf{X}$ / $\mathbf{J}$ (ADMM-LI / baseline)
\begin{algorithmic}[1]
\For{$r=1:R$}
\State $\nu_n^{(r)} \gets (\epsilon_0 + ||\mathbf{x}_n||_2)^{-1}$ / $(\epsilon_0 + ||\mathbf{j}_n||_2)^{-1}$
\For{$s=1:S_\text{max}$}
\State $\mathbf{Z}^{(r,s+1)}\gets$ \eqref{eq:Z_info_final} / \eqref{eq:Z_noinfo_final}
\State $\mathbf{X}^{(r,s+1)}/\mathbf{J}^{(r,s+1)}\gets$ \eqref{eq:X_info_final} / \eqref{eq:j_noinfo_final}
\State $\mathbf{W}^{(r,s+1)}\gets$ \eqref{eq:W_info_final} / \eqref{eq:W_noinfo_final}
\EndFor
\EndFor
\State $\mathbf{X}$ / $\mathbf{J} \gets \mathbf{X}^{(R_\text{max},S_\text{max})}$ / $\mathbf{J}^{(R_\text{max},S_\text{max})}$
\end{algorithmic}
\end{algorithm}
\subsection{Computational complexity and implementation}
In both \gls{ADMM-LI} and the baseline, the operations with the higher cost are the matrix inversion and matrix multiplications for determining $\mathbf{Z}^{(s+1)}$, \eqref{eq:Z_info_final} and \eqref{eq:Z_noinfo_final}, respectively. For \gls{ADMM-LI}, the matrix inversion has complexity $\bigO (N^6)$, while the inner multiplication has complexity $\bigO (NMT + N^2T)$, and the resulting multiplication of the inverted matrix by the vectorization is $\bigO(N^4)$, resulting in $\bigO(N^6+N^2T+NMT)=\bigO(N^6+NMT)$. On the other hand, for the baseline, the complexity is~$\bigO (N^3+N^2+NMT)=\bigO(N^3+NMT)$. Although the complexity is high, some properties of the matrices in question can be explored to reduce computational time. First, we note that $\hat{\mathbf{H}}\hat{\mathbf{H}}\herm$ and $\boldsymbol{\Phi}\herm\boldsymbol{\Phi}$ are hermitian positive definite matrices, and so is their Kronecker product and its sum with the identity matrix. Thus the linear problems explicitly solved in \eqref{eq:Z_info_final} and \eqref{eq:Z_noinfo_final} can be more efficiently solved with numerical methods. In practice, the use of LU decomposition and forward and back substitution can significantly decrease the run-time of the algorithms, although their asymptotic complexity will remain the same.

\begin{table}[]
    \centering
    \begin{tabular}{l c c}
    \hline
    Parameter & Symbol & Value \\
    \hline
    Number of users & $N$ & 24 \\
    Number of active users & $K$ & 4 \\
    Number of antennas & $M$ & 32 \\
    Carrier frequency & $f$ & 1710~MHz \\
    Min. and max. distances of users to \gls{BS} & $r_{\min}$, $r_{\max}$ & 20~m, 80~m \\
    Min. and max. angles of users to X-axis & $\theta_{\min}$, $\theta_{\max}$ & $-3\pi/7$, $3\pi/7$ \\
    Pilot length & $T$ & 6~symbols \\
    \hline
    \end{tabular}
    \caption{Summary of the base parameters used throughout the simulations.}
    \label{tab:parameters}
\end{table}

\section{Results}\label{sec:results}
{In this section, we evaluate the performance of our proposal and compare it to the baseline under several different network configurations. In Table~{\ref{tab:parameters}}, we summarize the base parameters, which are the same in all simulations unless stated otherwise.} As our base configuration, we consider $N=24$ users and, at each \gls{TS}, users are active with probability $p=4/N$. The value for $N$ is chosen due to the complexity of the algorithms under study, but it can be representative of a network with more users grouped into clusters, operating orthogonally in the frequency, time and/or pilot sequence domains. The \gls{BS} is equipped with a linear antenna positioned along the Y-axis, with $M=32$ elements, each located at $\boldsymbol{\delta}_m=[0,(m-1)\lambda-(M-1)\lambda/2]$, and operates at carrier frequency $f=1710$~MHz, as used, in, e.g., NB-IoT {\mbox{\cite{3GPP}}}. {The minimum and maximum distance of users to the {\gls{BS}} are $r_{\min}=20$~m and $r_{\max}=80$~m, respectively. Here, the distances are chosen such that users stay in the radiative near-field for $M$ between $M=32$ and $M=64$, as this range will be considered in one of our experiments.} Lastly, the minimum and maximum angles of users to the X-axis are set to $\theta_{\min}=-\pi/2$ and $\theta_{\max}=\pi/2$, {occupying the first and fourth quadrants and facing the antenna array.} We let the pilot transmission take $T=6$~symbols, and each entry of the pilot matrix is drawn from a complex normal distribution, {with the whole pilot normalized}. 

We evaluate the approaches using the complement of the balanced accuracy as the main performance metric, defined as
\begin{equation}\label{eq:inac}
    1 - A = 1 - \frac{\text{TPR}+\text{TNR}}{2}.
\end{equation}
Here, $\text{TPR} \triangleq |\text{TP}|/K$ represents the true positive rate, where $|\text{TP}|$ is the number of correctly detected active users. Conversely, $\text{TNR} \triangleq |\text{TN}|/(N-K)$ is the true negative rate, where $|\text{TN}|$ denotes the number of correctly identified inactive users.

The hyper-parameters of the \gls{ADMM} algorithms are selected a priori such that the resulting performance is close to the best achievable performance based on training from $10^{4}$ iterations, varying for each method. The parameters for each approach are presented in Table \ref{tab:hyper}. 
To achieve reliable and unbiased results, each experiment runs for $10^6$ iterations, and the position of the users, as well as their pilot sequences, are randomly sampled at each Monte Carlo iteration.

\begin{table}[]
    \caption{\gls{ADMM} hyper-parameters for the methods with known and unknown users' position.}
    \centering
    \begin{tabular}{c c}
    \hline
    Parameter & value \\
    \hline
    $\beta$         & $10^{-5}$\\
    $\rho$          & $10^{-1}$\\
    $\epsilon_0$    & $10^{-1}$\\
    $R$       & $10$ \\
    $S$       & $10$ \\
    \hline
    \end{tabular}
    \label{tab:hyper}
\end{table}

{\subsection{Impact of the \gls{SNR}}
We evaluate the performance of \gls{ADMM-LI} as the \gls{SNR} varies from $-10$ to $5$~dB and present the results in Fig. \ref{fig:SNR} for $M=16$ and $M=32$ antennas. While the performance of both the baseline and \gls{ADMM-LI} increases with the SNR, the performance gap is much more evident. Already at SNR = $0$~dB, \gls{ADMM-LI} outperforms the baseline by more than one order of magnitude with $M=16$ (dashed lines), and almost two orders of magnitude with $M=32$ (continuous lines). At  SNR = $5$~dB, the gap is of more than one and two orders of magnitude for $M=16$ and $M=32$, respectively. However, at lower \gls{SNR} values, the signal is so corrupted by noise that the received signal does not resemble the combination of pilot sequence and channel response as much, resulting in modest performance improvements. 

\begin{figure}
    \centering
    \includegraphics[width=\linewidth]{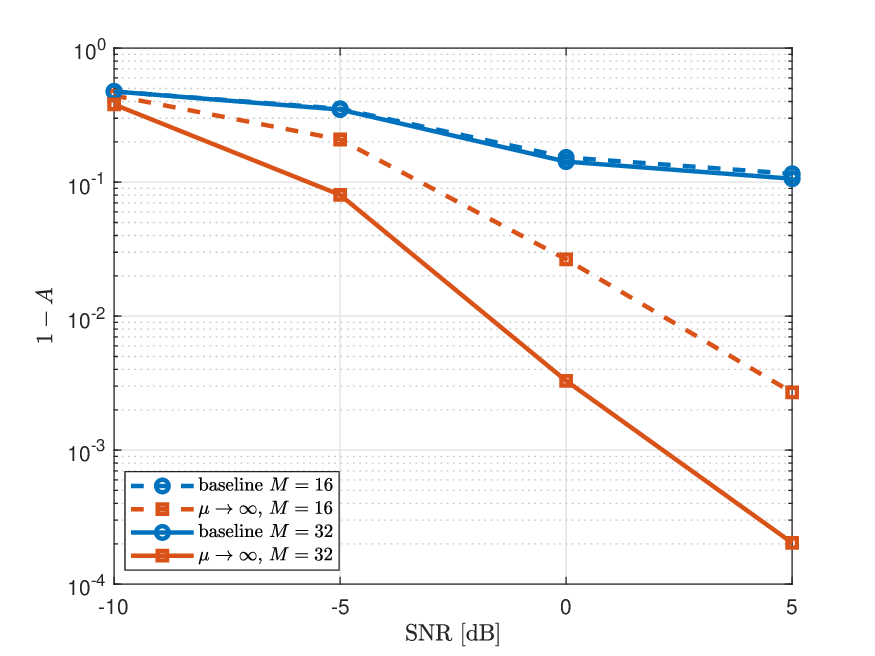}
    \caption{Performance comparison between the baseline and the approach with location information as a function of the SNR. }
    \label{fig:SNR}
\end{figure}}

\subsection{Multipath}
\begin{figure}
    \centering
    \includegraphics[width=\linewidth]{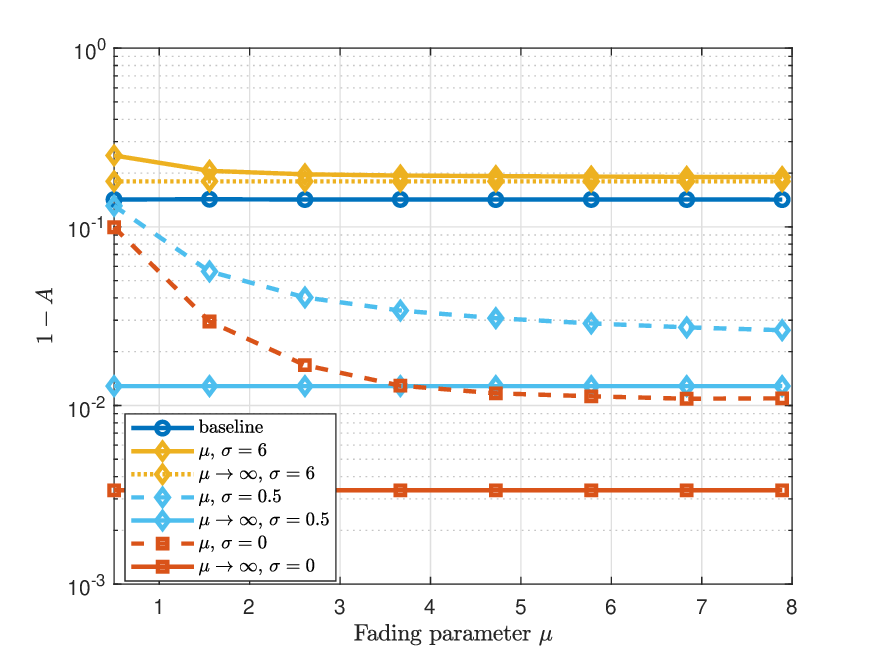}
    \caption{Performance comparison between the baseline and the approaches with perfect and imperfect location information for the case with SNR = $0$~dB as a function of the fading parameter $\mu$.}
    \label{fig:multipath}
\end{figure}

To evaluate the impact of multipath in the performance of our proposal, we vary the Rician fading parameter from $\mu=0.5$, meaning a very weak \gls{LoS}, to a dominant \gls{LoS} component with $\mu\approx8$, and present the results in Fig. \ref{fig:multipath}, {where the curves indicated with $\mu\to\infty$ stand for the case with only {\gls{LoS}}.} In the cases with weak \gls{LoS}, the location of the users provides little information on the resulting channel, but it is enough to improve the performance over the baseline. {This is true for the cases with small or inexistent errors, the dashed orange ($\sigma=0$) and dashed cyan ($\sigma=0.5$) curves in Fig.~{\ref{fig:multipath}}, which already at $\mu\approx1.5$ provide a performance gain of almost one order of magnitude compared to the  baseline (blue continuous curve).} {As the \hbox{\gls{LoS}} becomes stronger, the performance improvement is more significant, and the curves where $\mu$ is varied approach their full \hbox{\gls{LoS}} counterparts.} Interestingly, as the errors in the location estimate increase, the multipath has less impact on the performance, with the curves converging faster. 

{\subsection{Imperfect location estimate}
While the proposed method greatly outperforms the baseline, assuming perfect location information is too optimistic in most cases. This is possible if users are completely static and if their positions are precisely and explicitly informed to the \gls{BS}, which is not always the case. However, it is possible to estimate the positions of users based on the received signal using, e.g., the MUSIC algorithm \cite{Haghshenas:arxiv:2025,Gurgunoglu:IWCL:2025} or Qiao's~\cite{Qiao:2024:ITC}~\etal~proposal, although with some errors. To model these possible errors, we vary $\sigma$ from $0$~m (perfect location estimation) to $10$~m and present the result in Fig. \ref{fig:imperfectSNR0} for SNR = $0$~dB. The results in Fig. \ref{fig:imperfectSNR0} show an interesting effect: while the performance of \gls{ADMM-LI} decreases as errors are introduced in the location estimate, this method can still be used as long as the error is not too high, provided that the \gls{LoS} component is strong enough. {This is evident from the purple and green curves, which shows the cases with strong {\gls{LoS}} component ($\mu=5$ and $\mu\to\infty$, respectively), outperforming the baseline at up to $\sigma=4$. On the other hand, if the location information has high errors and is not reliable, it is better to use the baseline instead, as it has better performance in such cases, and has much lower complexity.} However, small errors of around 1.5~m, even for SNR = $0$~dB are reasonable to assume, as Haghshenas \etal \cite{Haghshenas:arxiv:2025} observerd when they evaluated MUSIC's performance in a setup similar to ours. On top of this, tracking algorithms \cite{Guerra:2021:ITSP} can further improve the estimates. 
\begin{figure}
    \centering
    \includegraphics[width=\linewidth]{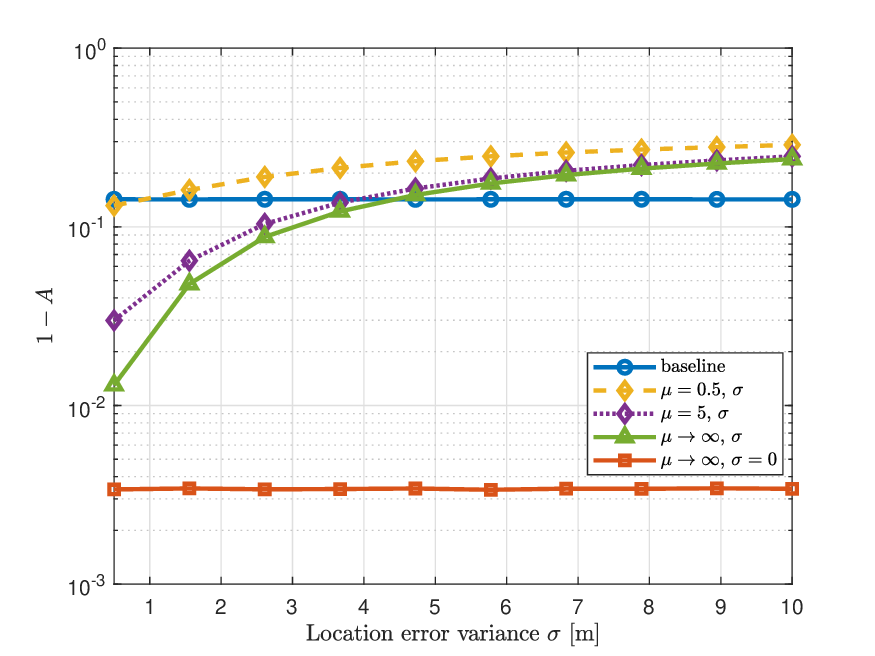}
    \caption{Performance comparison between the baseline and the approaches with perfect and imperfect location information for the case with SNR = $0$~dB as a function of the position error variance.}
    \label{fig:imperfectSNR0}
\end{figure}}

\subsection{Varying pilot length}
\begin{figure}
    \centering
    \includegraphics[width=\linewidth]{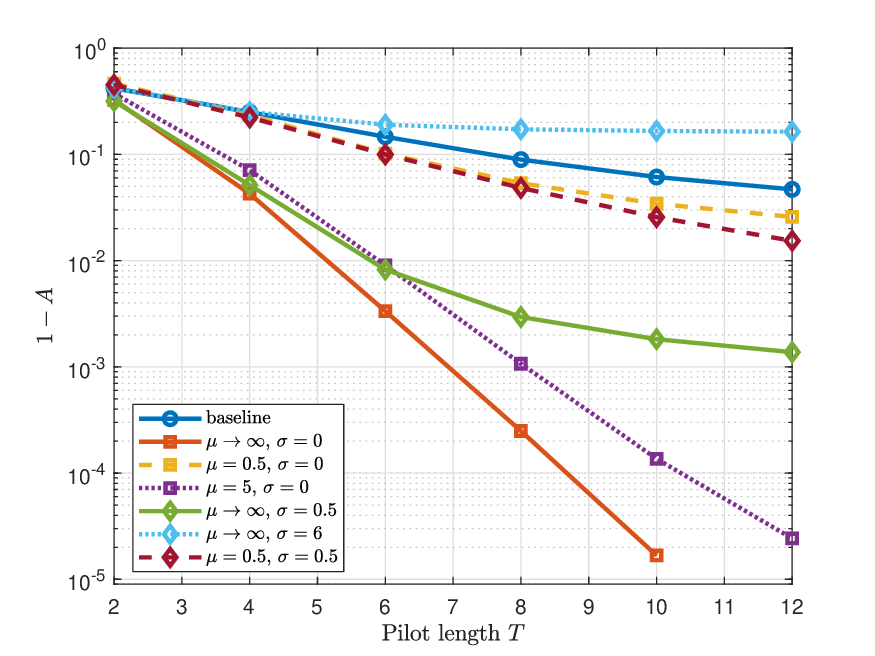}
    \caption{Performance comparison between the baseline and the approaches with perfect and imperfect location information for the case with SNR = $0$~dB as a function of the pilot length.}
    \label{fig:pilotLength}
\end{figure}

We evaluate the performance of \gls{ADMM-LI} as the pilot length varies from $T=2$ to $T=12$~symbols and present the results in Fig. \ref{fig:pilotLength}. Alongside the baseline and the case with perfect location information, we present results for $\sigma=\{0.5, 6\}$ and $\mu=\{0.5, 5\}$. When the location errors are large {(cyan curve, $\sigma=6$)}, the proposal slightly outperforms the baseline only in shorter pilot lengths, with their performances matching at $T=4$, and the baseline outperforming it for $T>4$. As the pilot length increases, errors in the location estimate are detrimental to the performance, as the introduced error in the channel matrix results in low correlation with the received signal. In this case, even if there is only \gls{LoS}, considering only the pilot length is more beneficial to \gls{AUD}. If location errors are small, even with a weak \gls{LoS} the proposal outperforms the baseline, {as indicated by the yellow ($\mu=0.5$, $\sigma=0$), purple ($\mu=5$, $\sigma=0$), green ($\mu\to\infty$, $\sigma=0.5$) and dark red curves ($\mu=0.5$, $\sigma=0.5$).} Moreover, even small errors in the location of users, {the green curve}, can limit the performance of the proposal with diminishing returns as the pilot length increases, although still significantly outperforming the baseline.

\subsection{Number of users}
\begin{figure}
    \centering
    \includegraphics[width=\linewidth]{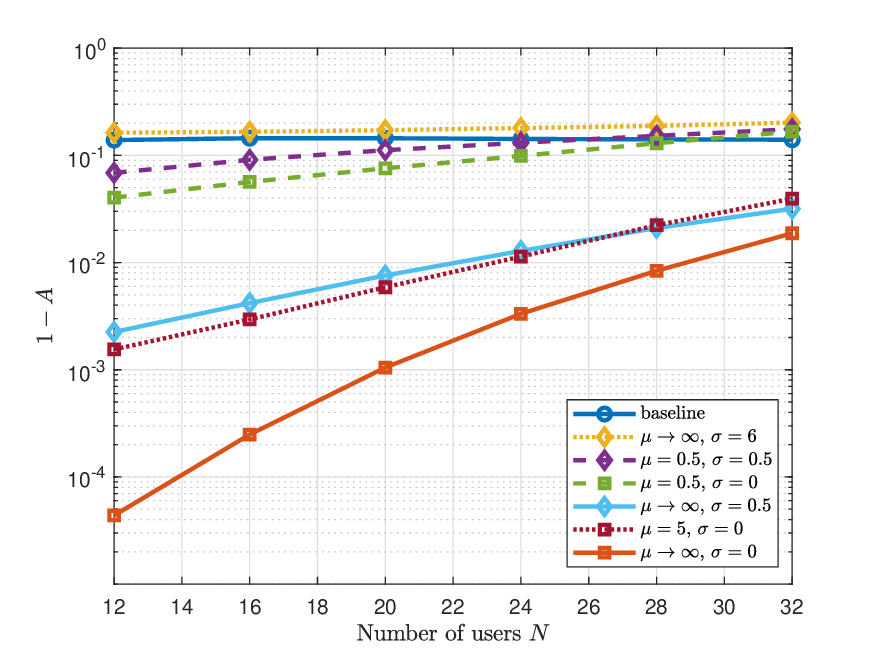}
    \caption{Performance comparison between the baseline and the approaches with perfect and imperfect location information for the case with SNR = $0$~dB as a function of the number of users.}
    \label{fig:numberOfUsers}
\end{figure}

We vary the number of users from $N=12$ to $N=32$, while the remaining parameters stay fixed, and present the results in Fig. \ref{fig:numberOfUsers}. The proposed method performs exceptionally well in lower counts of users, outperforming the baseline by more than four orders of magnitude. {While \hbox{\gls{ADMM-LI}} without location errors and strong \hbox{\gls{LoS}} always outperforms the other approaches, it is much more sensitive to the number of users, degrading with a higher rate than the case without errors, and even the baseline.}  Moreover, as the number of users increases, the location errors have less impact in the final performance. While it does not prevent performance from degrading, using the proposal even with errors is preferred over the baseline, as long as they are small. In this case, having some information of the channel, is beneficial to the \gls{AUD}, as it helps to further differentiate between users compared to only when the pilot sequence is used, even if the location is not entirely correct, {as is the case of the dark red ($\mu=0.5$, $\sigma=0.5$), yellow ($\mu=0.5$, $\sigma=0$) and green  ($\mu\to\infty$, $\sigma=0.5$) curves}. 

\subsection{Number of active users}
\begin{figure}
    \centering
    \includegraphics[width=\linewidth]{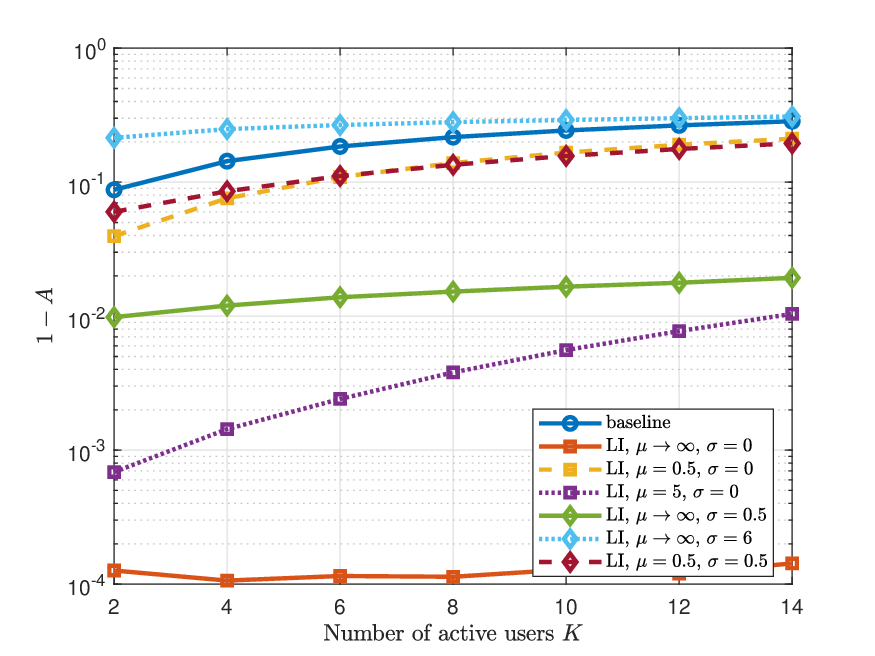}
    \caption{Performance comparison between the baseline and the approaches with perfect and imperfect location information for the case with SNR = $0$~dB as a function of the number of active users.}
    \label{fig:numberOfActiveUsers}
\end{figure}

{We vary the number of active users from $K=2$ to $K=14$ and present the results in Fig. \ref{fig:numberOfActiveUsers}. As it is expected in the \gls{AUD}, as the number of active users increase, the accuracy decreases. In ideal conditions ($\sigma=0$ and $\mu\to\infty$), \gls{ADMM-LI} is the least sensitive, still fairly outperforming the baseline at high values of $K$. However, even in the presence of weak \gls{NLoS} ($\mu=5$), the performance degrades significantly with $K$, although still outperforming the baseline. When errors in the location are present ($\sigma\neq 0$), the performance degrades slower with $K$.}

\section{Conclusions}\label{sec:conclusion}
In this paper we explored one way to use the position of users to solve the \gls{AUD} problem. Specifically working in the near-field, we proposed a method that takes into account prior knowledge of the \gls{BS} about the users' localization and the multipath statistics. The proposed method greatly outperformed the standard approach, though with increased computation complexity of $\mathcal{O}(N^6+NMT)$ compared to $\mathcal{O}(N^3+NMT)$, and the requirement of perfect knowledge of the users' position. However, we showed that our method can be used with imperfect localization estimation, still with a significant performance improvement when the position error is not too high and the \gls{LoS} component is dominant.

\appendix\label{ap:formulationADMMLI}
We minimize \eqref{eq:Lagrangian} by taking the gradient with respect to each variable at a time and setting it to $\mathbf{0}$. For $\mathbf{Z}$, we first define the functions 
\begin{equation*}
    g_1(\mathbf{Z}) \triangleq \frac{1}{2}||\boldsymbol{\Phi}\mathbf{Z}\hat{\mathbf{H}} - \mathbf{Y}||_F^2,\quad g_2(\mathbf{Z})=\frac{\rho}{2}||\mathbf{X}^{(s)}-\mathbf{Z}+{\mathbf{W}^{(s)}}/{\rho}||_F^2,
\end{equation*}
\begin{equation*}
g_3(\mathbf{Z}) \triangleq \frac{1}{2}\tr((\boldsymbol{\Phi}\mathbf{Z}\mathbf{H})(\boldsymbol{\Phi}\mathbf{Z}\mathbf{H})\herm, \quad g_4(\mathbf{Z})\triangleq\tr((\boldsymbol{\Phi}\mathbf{Z}\hat{\mathbf{H}})\mathbf{Y}\herm).
\end{equation*}
 We then take the gradient of  \eqref{eq:Lagrangian} with respect to $\mathbf{Z}$, and set it to $\mathbf{0}$, as
\begin{align*}\label{eq:Z_info_re} 
\nabla_\mathbf{Z} \mathcal{L}(\mathbf{X},\mathbf{Z}; \mathbf{W}) = &\nabla_\mathbf{Z} (g_1(\mathbf{Z}) + g_2(\mathbf{Z})) \\= & \nabla_\mathbf{Z} g_1(\mathbf{Z}) + \nabla_\mathbf{Z}g_2(\mathbf{Z}) = \mathbf{0}.\numberthis
\end{align*}

Expanding on $\nabla_\mathbf{Z}g_1(\mathbf{Z})$, we have
\begin{equation}\label{eq:nablaZ1}
\begin{split}
     &\nabla_\mathbf{Z}g_1(\mathbf{Z}) = \nabla_\mathbf{Z} \left(\frac{1}{2}||\boldsymbol{\Phi}\mathbf{Z}\hat{\mathbf{H}} - \mathbf{Y}||_F^2\right)  \\ &=\frac{1}{2}\nabla_\mathbf{Z} \left(\tr((\boldsymbol{\Phi}\mathbf{Z}\hat{\mathbf{H}} - \mathbf{Y})(\boldsymbol{\Phi}\mathbf{Z}\hat{\mathbf{H}} - \mathbf{Y})\herm\right)\\
    & = \frac{1}{2}\nabla_\mathbf{Z} \left(\tr((\boldsymbol{\Phi}\mathbf{Z}\mathbf{H})(\boldsymbol{\Phi}\mathbf{Z}\mathbf{H})\herm\right)-\nabla_\mathbf{Z}(\tr((\boldsymbol{\Phi}\mathbf{Z}\mathbf{H})\mathbf{Y}\herm)) \\
    & = \frac{1}{2}(\nabla_\mathbf{Z}g_3(\mathbf{Z})-\nabla_\mathbf{Z}g_4(\mathbf{Z})).
\end{split}
\end{equation}
Then, $\nabla_\mathbf{Z} g_3(\mathbf{Z})$ is expanded as
\begin{equation}\label{eq:nablaZ1:1}
    \begin{split}
        &\nabla_\mathbf{Z} g_3(\mathbf{Z}) = \frac{1}{2}\left(\tr((\boldsymbol{\Phi}\mathbf{Z}\mathbf{H})(\boldsymbol{\Phi}\mathbf{Z}\mathbf{H})\herm\right)  \\ & =  \frac{1}{2}\nabla_\mathbf{Z} \left(\tr(\boldsymbol{\Phi}\herm\boldsymbol{\Phi}\mathbf{Z}\hat{\mathbf{H}}\hat{\mathbf{H}}\herm\mathbf{Z}\right)\\
        & = \frac{1}{2}\boldsymbol{\Phi}\herm\boldsymbol{\Phi}\mathbf{Z}\hat{\mathbf{H}}\hat{\mathbf{H}}\herm + \frac{1}{2}(\boldsymbol{\Phi}\herm\boldsymbol{\Phi})\herm\mathbf{Z}(\hat{\mathbf{H}}\hat{\mathbf{H}}\herm)\herm \\
        & = \boldsymbol{\Phi}\herm\boldsymbol{\Phi}\mathbf{Z}\hat{\mathbf{H}}\hat{\mathbf{H}}\herm,
    \end{split}
\end{equation}
and for $\nabla_\mathbf{Z} g_4(\mathbf{Z})$ we have
\begin{equation}\label{eq:nablaZ1:2}
    \begin{split}
        &\nabla_\mathbf{Z} g_4(\mathbf{Z}) = \nabla_\mathbf{Z}(\tr((\boldsymbol{\Phi}\mathbf{Z}\hat{\mathbf{H}})\mathbf{Y}\herm)) = \nabla_\mathbf{Z}(\tr(\hat{\mathbf{H}}\mathbf{Y}\herm\boldsymbol{\Phi}\mathbf{Z})) \\ & = (\hat{\mathbf{H}}\mathbf{Y}\herm\boldsymbol{\Phi})\herm = \hat{\mathbf{H}}\herm\mathbf{Y}\boldsymbol{\Phi}\herm.
    \end{split}
\end{equation}
Substituting \eqref{eq:nablaZ1:2} and \eqref{eq:nablaZ1:1} in \eqref{eq:nablaZ1} we obtain
\begin{equation}\label{eq:nablaZ1:solved}
    \nabla_\mathbf{Z} \left(\frac{1}{2}||\boldsymbol{\Phi}\mathbf{Z}\hat{\mathbf{H}} - \mathbf{Y}||_F^2\right) = \boldsymbol{\Phi}\herm\boldsymbol{\Phi}\mathbf{Z}\hat{\mathbf{H}}\hat{\mathbf{H}}\herm + \hat{\mathbf{H}}\herm\mathbf{Y}\boldsymbol{\Phi}\herm.
\end{equation}
Now, $\nabla_{\mathbf{Z}}g_2(\mathbf{Z})$ is solved as
\begin{align*}
     \nabla_{\mathbf{Z}}g_2(\mathbf{Z})& =\nabla_\mathbf{Z}\left(\frac{\rho}{2}||\mathbf{X}^{(s)}-\mathbf{Z}+{\mathbf{W}^{(s)}}/{\rho}||_F^2\right) \\&= -\rho (\mathbf{X}^{(s)}-\mathbf{Z}+{\mathbf{W}^{(s)}}/{\rho}).\numberthis\label{eq:nablaZ2:solved}
\end{align*}
Finally, with \eqref{eq:nablaZ1:solved} and \eqref{eq:nablaZ2:solved} we isolate $\mathbf{Z}$ as
\begin{equation}\label{eq:SolveZ}
    \boldsymbol{\Phi}\herm\boldsymbol{\Phi}\mathbf{Z}\hat{\mathbf{H}}\hat{\mathbf{H}}\herm + \rho\mathbf{Z} = \hat{\mathbf{H}}\herm\mathbf{Y}\boldsymbol{\Phi}\herm + \rho \mathbf{X}^{(s)} + {\mathbf{W}^{(s)}}
\end{equation}
 \begin{align*}
 \vectorial(\mathbf{Z}) &=  (\hat{\mathbf{H}}\hat{\mathbf{H}}\herm\otimes \boldsymbol{\Phi}\herm\boldsymbol{\Phi} + \rho \mathbf{I}_{N^2})^{-1}\\&\vectorial(\hat{\mathbf{H}}\herm\mathbf{Y}\boldsymbol{\Phi}\herm + \rho \mathbf{X}^{(s)} + {\mathbf{W}^{(s)}}),\numberthis
 \end{align*} 
where $\otimes$ denotes the Kroenecker product, and $\vectorial({\cdot})$ is the vectorization operation. Then, ${\mathbf{Z}^{(s+1)}}$ is obtained by properly reshaping $\vectorial(\mathbf{Z})$.

On the other hand, for $\mathbf{X}$, we have
\begin{align*}
    &\nabla_\mathbf{X}\mathcal{L}(\mathbf{X},\mathbf{Z};\mathbf{W}) =\\ & \nabla_\mathbf{X}\left(\beta \sum_{n=1}^N \nu_n^{(r)}||\mathbf{x}_n||_2 + \frac{\rho}{2}||\mathbf{X}-\mathbf{Z}^{(s+1)}+{\mathbf{W}^{(s)}}/{\rho}||_F^2\right).
\end{align*}
To solve for $\mathbf{X}$, we explore the fact that $\mathbf{X}$ should only have non-zero elements in its diagonal, simplifying to
\begin{align*}
    \nabla_\mathbf{X}\left(\beta \sum_{n=1}^N \nu_n^{(r)}||X_{n,n}||_2 + \frac{\rho}{2}||X_{n,n}-Z_{n,n}^{(s+1)}+{W_{n,n}^{(s)}}/{\rho}||_2^2\right).\numberthis
\end{align*}
By defining ${d}_n = {Z}_{n,n}^{(s+1)}-{W_{n,n}^{(s)}}/{\rho}$, $\mathbf{X}$ can be solved using the iterative shrinkage thresholding approach \cite{Qin:MPC:2013} as
\begin{equation}
    X_{n,n} = \frac{d_n}{|d_n|} \max\left(0, |d_n| - \beta\frac{ \nu_n}{\rho}\right).
\end{equation}

\bibliographystyle{IEEEtran}
\bibliography{main}
\end{document}